%% file: main.tex
\documentclass[conference]{IEEEtran}
\IEEEoverridecommandlockouts
\usepackage{cite}
\usepackage{amsmath,amssymb,amsfonts}
\usepackage{algorithmic}
\usepackage{graphicx}
\usepackage{textcomp}
\usepackage{xcolor}
\def\BibTeX{{\rm B\kern-.05em{\sc i\kern-.025em b}\kern-.08em
    T\kern-.1667em\lower.7ex\hbox{E}\kern-.125emX}}
    
\usepackage{xspace}
\usepackage{multirow}
\usepackage{subcaption}
\usepackage[hidelinks]{hyperref}
\usepackage{xurl}

\usepackage{booktabs}
\usepackage{tcolorbox} 
\tcbset{colback=white,arc=0mm}
\usepackage{color, colortbl}

\newcommand{\hopsworks}{Hopsworks\xspace}

\begin{document}

\title{Cloud-native RStudio on Kubernetes for \hopsworks\\
}

\author{
\IEEEauthorblockN{{
    Gibson~Chikafa\IEEEauthorrefmark{1},
    Sina~Sheikholeslami\IEEEauthorrefmark{2},
    Salman~Niazi\IEEEauthorrefmark{1},
	Jim~Dowling\IEEEauthorrefmark{1}\IEEEauthorrefmark{2},
	Vladimir~Vlassov\IEEEauthorrefmark{2}}}
	
\IEEEauthorblockA{
\IEEEauthorrefmark{1}Hopsworks, Stockholm, Sweden\\
Email: \{gibson, salman, jim\}@logicalclocks.com\\
\IEEEauthorrefmark{2}KTH Royal Institute of Technology, Stockholm, Sweden\\
Email: \{sinash, jdowling, vladv\}@kth.se}
}

\maketitle

\begin{abstract}
In order to fully benefit from cloud computing, services are designed following the ``multi-tenant'' architectural model, which is aimed at maximizing resource sharing among users. However, multi-tenancy introduces challenges of security, performance isolation, scaling, and customization. RStudio server is an open-source Integrated Development Environment (IDE) accessible over a web browser for the R programming language. We present the design and implementation of a multi-user distributed system on Hopsworks, a data-intensive AI platform, following the multi-tenant model that provides RStudio as Software as a Service (SaaS). We use the most popular cloud-native technologies: Docker and Kubernetes, to solve the problems of performance isolation, security, and scaling that are present in a multi-tenant environment. We further enable secure data sharing in RStudio server instances to provide data privacy and allow collaboration among RStudio users. We integrate our system with Apache Spark, which can scale and handle Big Data processing workloads. Also, we provide a UI where users can provide custom configurations and have full control of their own RStudio server instances. Our system was tested on a Google Cloud Platform cluster with four worker nodes, each with 30GB of RAM allocated to them. The tests on this cluster showed that 44 RStudio servers, each with 2GB of RAM, can be run concurrently. Our system can scale out to potentially support hundreds of concurrently running RStudio servers by adding more resources (CPUs and RAM) to the cluster or system. 
\end{abstract}


\input{sections/introduction}

\input{sections/background}
\input{sections/design}
\input{sections/discussion}
\input{sections/evaluation-results}
\input{sections/conclusion}
\bibliographystyle{IEEEtran}
\typeout{}
\bibliography{bibliography}
\end{document}

%% file: sections/introduction.tex
\section{Introduction}
One of the important developments that have led to the success of data science is the development of systems that offer an interactive command-line interface in which code can be run immediately without having to go through the traditional edit/compile/execute cycle~\cite{4160251}. An interactive environment lets users look at data, test new ideas, combine algorithmic approaches, and evaluate their outcomes directly. This is one of the reasons that programming languages such as Python, R, MATLAB, SQL, and Julia have become very popular among data scientists~\cite{top_data_science_languages}.

In recent years, efforts have been made to develop multi-user platforms to support interactive data analytics. One of the most popular such platforms today is JupyterHub~\cite{jupyterhub_architecture}, an open-source service that creates on-demand Jupyter Notebook servers. JupyterHub enables multi-tenancy by giving several users access to the same computational environments and resources while eliminating the burden of installation and maintenance tasks. Some systems developed on top of JupyterHub have been described in~\cite{9289692} and~\cite{9323748}. It can be argued that JupyterHub is one of the reasons Python is more popular to organizations, as it gives easy access to Jupyter Notebook servers on-demand in a multi-user environment. 

For RStudio, there are no open-source systems to our knowledge that provide the same capabilities as JupyterHub does for Python, and the few cloud-based solutions available are not open-source. RStudio Cloud \footnote{https://rstudio.cloud/} is the lightweight, cloud-based solution that offers the RStudio IDE in the cloud. There is a free limited version for individual use, but the features necessary for group usage (e.g., classroom, organization) require a paid license. 

In this paper, we present the integration of RStudio (community version)  on Hopsworks, an open-source enterprise platform for scale-out data science, following the multi-tenant model. We use the leading cloud-native technologies (Docker and Kubernetes) to solve the challenges that are introduced in a multi-tenant environment, such as performance isolation, security, and scaling. We also enable secure data sharing in RStudio server instances through the introduction of HopsFS-mount, which allows to mount HopsFS (a highly scalable version of HDFS) in the RStudio server instances. Hopsworks provides several other features for big data engineering and machine learning, such as Apache Spark and Hadoop, so with this integration, such services will become seamlessly available to RStudio users on the platform. With our contribution to providing the multi-user environment for RStudio on open-source Hopsworks while eliminating installation and maintenance tasks, we are promoting collaboration among data scientists and the learning and teaching of R.

%% file: sections/background.tex
\section{Background}

Cloud-based multi-user environments for RStudio are required by organizations such as universities that have courses on data science or machine learning with R and data science teams in companies. To leverage economies of scale, such environments should be designed in a way that the underlying system resources are shared and utilized efficiently. For universities, the number of users can potentially grow to hundreds; therefore, the system should be designed with proper isolation and scaling mechanisms. On the other hand, companies would like to store big data, share data securely and handle big data processing workloads; hence the system should have capabilities of big data storage, scaling, secure data sharing and
integration with big data processing frameworks.

\subsection{Multi-tenancy}
To fully benefit from cloud computing, services should be designed following the “multi-tenant” architectural model, which is aimed at maximizing resource sharing among users~\cite{8090268}. Generally, multi-tenancy can be described as a model that allows all information system resources to be shared by multiple users and sub-users at the same time. The benefits of multi-tenancy include maximum resource utilization, separation of hardware from software failure, reduced costs, reduced power consumption, saving money, less carbon emission, less over-provisioning, reduced costs, and increasing profit for the cloud vendor\cite{6830928}. All these benefits stem from either virtualization, resource sharing, or combining both of them; thus, multi-tenancy can also be expressed as: “Multi-Tenancy = virtualization + resource sharing”. In particular, the separation of the hardware failure from the software failure is achieved by virtualization. On the other hand, resource sharing will increase utilization which will consequently lead to a reduction in costs and energy usage by making the resource available to more than one customer.

However, there are also some issues and challenges that multi-tenant environments instigate. These challenges include performance isolation, security, and scaling.
\subsubsection{Multi-tenancy and Performance Isolation}
Matei Zaharia, co-founder of Databricks \cite{databricks}, in his talk titled ``Lessons from Large-Scale Cloud Software at Databricks'' states that performance isolation is one of the critical challenges when building multi-tenant services \cite{matei_zaharia_talk}, but it is not studied in research in the academia. Performance isolation can be defined as a guarantee that the performance and resource consumption of one tenant does not adversely affect other tenants \cite{Ochei2015EvaluatingDO}. Performance isolation between the different tenants is hard to achieve in SaaS due to the high level of  resource sharing~\cite{2405178.2405184}. If resources are not properly isolated, it may lead users who demand resources within their limits to be affected by those who exceed their limits. As an example, Databricks has reported that 20\% of significant outages reported by its users are caused by insufficient user isolation \cite{matei_zaharia_talk}. This situation can lead to a loss of trust in a service or platform, which can eventually lead to a business-critical situation. Thus, handling the performance problem in a fair way is as important as providing the functional services of the application. There are two objectives in performance isolation: (i) preventing one tenant from adversely affecting the performance of other tenants, and (ii) ensuring that the performance for the different tenants complies with their Service Level Agreements \cite{4285271}.

A simple solution to performance isolation would be to give each user their own computational resources, e.g., servers. In this case, the interference of different users is avoided using isolated resource sharing. However, this is not a cost-effective solution since several resources can become unnecessarily idle.

The majority of existing solutions achieve performance isolation between different tenants by assigning separate virtual machines. However, these approaches inherently result in less operational cost benefits in comparison to application-level multi-tenancy. Another feasible solution is to guarantee the performance demands of multiple users by relying on the elastic scalability of the system \cite{TEKINERDOGAN2017127}. With elasticity, the system can scale itself up or down according to the minimum demands of each tenant.
\subsubsection{Multi-tenancy and Security}
Since the inception of cloud computing, security has always been a serious concern. This is due to the very nature of multi-tenancy; allowing several users to share the same hardware allows both the attacker and the victim to share the same physical device \cite{6830928}. Multi-tenancy has been recognized as a security loophole in cloud computing by several researchers. The authors of \cite{Subashini2011ASO}, who conducted a survey on security issues in service delivery models in clouds, state that multi-tenancy is a major feature of cloud computing that may lead to confidentiality violation. \cite{5557959} outlines the best practices for building secure clouds but yet clearly recognizes multi-tenancy and shared technology issues as security challenges for a cloud environment. In a report released by the Cloud Security Alliance (CSA) titled ``Security as a Service'' \cite{5708519}, they state that multi-tenancy introduces new targets for intrusion and less assurance for data segregation. Furthermore, \cite{monitoring_virtual_private_network} emphasizes the
fact that multi-tenancy may enable information leakage and
increase attack options that affect the security of the clouds. Also, \cite{it_audit_to_essure_secure_cloud_computing}, \cite{security_audits_of_virtual_infra} recognize  multi-tenancy among the serious issues in cloud security.

Security risks are present in multi-tenant environments regardless of the delivery model, i.e., at all levels of the deployment model, there are security issues and risks involved. Just as the capabilities are inherited in the delivery models, so are the security risks and issues \cite{Subashini2011ASO}. For instance, if the security issues are not handled well by the IaaS provider, they are propagated to PaaS and SaaS. Some of the fundamental security challenges in a multi-tenant environment are data storage security, data transmission security, application security, and security related to third-party resources \cite{Subashini2011ASO}.
\subsubsection{Multi-tenancy and Scaling}
 Scaling is defined as the process of adding or removing compute, storage, and network services to the system to meet the workload demands for resources in order to maintain system availability and performance as utilization increases \cite{6196111}. There are generally two scaling methods of a software system: vertical scaling (scale up) and horizontal scaling (scale out). Vertical scaling means supplementing the system with, e.g., more computing resources, more memory, higher disk bandwidth, and larger disk space. Horizontal scaling means splitting the workload or running the application across multiple machines, i.e., distributed, with similar conﬁgurations.
 
In a multi-tenant environment, the number of users can dynamically change. In order to support this dynamically increasing demand for resources from multi-tenants, the system should be able to duplicate computing resources to cope with the fluctuation of requests from tenants \cite{5704303}. 
\subsection{Cloud-Native Applications}
Cloud-native application development is an approach to building, running, and improving existing applications using the techniques and technologies suited for cloud computing. A cloud-native application consists of a collection of reusable and discrete components designed to run on all cloud environments: private, public, and hybrid clouds. Enabling technologies for cloud-native applications include containers, microservices, serverless functions, and immutable infrastructure deployed via declarative.

\textit{Containerization}, also known as  \textit{OS-level virtualization}, is a technology in which the OS kernel supports multiple isolated user-space environments or ~\cite{7036275} where processes within a user-space are confined to that user-space and do not have an overall view of the system. In such systems, the isolated environments are called containers. While each VM has its own OS, all containers share the same OS, which makes them lightweight. Containers are thus more lightweight and are faster to provision than VMs hence they are more efficient in the utilization of server resources. The three leading container technologies are Docker, LXD, and Rkt. Docker is arguably the most mature open-source project that provides a systematic way to automate the faster deployment of Linux applications inside portable containers. Docker extends the Linux Containers (LXC)\cite{linux_containers} with a kernel and application-level API that together run processes in isolation with isolated system resources such as CPU, memory, I/O, network, etc. In Docker, the first line of security isolation is achieved through Linux namespaces~\cite{linux_containers}, control groups or cgroups~\cite{linux_capabilities}, and Linux capabilities~\cite{linux_capabilities}. To harden the isolation, Linux Security Modules (LSMs) such as AppArmor and Security-Enhanced Linux (SELinux) may be used \cite{securing_linux_containers}. Furthermore, the Linux kernel Secure Computing mode (seccomp) may be used to limit system calls containers can make. To achieve performance isolation with Docker, it is possible to limit the amount of physical resources a container can consume. This is achieved through Linux cgroups. For example, using the \textit{cpu-quota} and \textit{memory} flags during container launch can limit the amount of CPU resources and memory a container can have, respectively. These limits help to prevent one tenant from taking all system resources and affecting other tenants.

One of the most popular open-source projects for managing containers across multiple hosts in a PaaS system is Kubernetes. Also known as K8s, Kubernetes provides mechanisms for automating deployment, scaling, and management of containerized applications~\cite{kubernetes_defn}. This automation is commonly referred to as container orchestration; thus, Kubernetes is a container orchestration framework.  Kubernetes takes care of the network provisioning, storage orchestration, health-checking container state with self-healing capabilities, and scaling for application containers within the cluster. This makes Kubernetes very suitable to run resilient distributed applications. 

\section{\hopsworks Data-intensive AI Platform}

\begin{figure}[!ht]
  \begin{center}
    \includegraphics[width=\columnwidth]{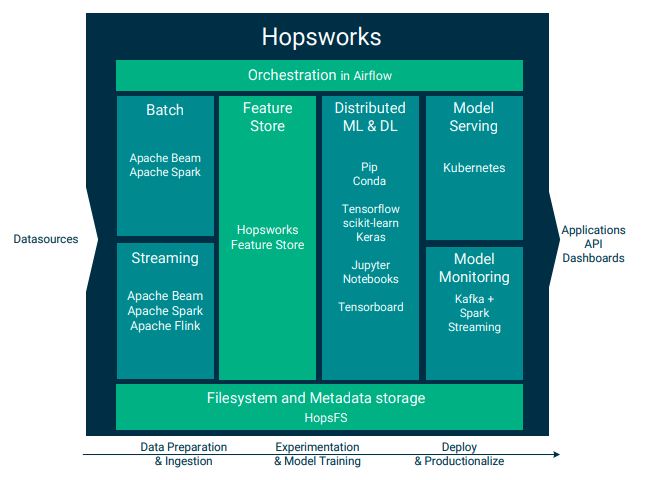}
  \end{center}
  \caption{Hopsworks Architecture~\cite{hopsworks_whitepaper}}
  \label{fig:hopsworks}
\end{figure}

\hopsworks is a full-stack open-source enterprise platform for scale-out data science, with support for GPU workloads and Big Data frameworks. It provides managed support for open-source frameworks for data engineering and data science through a unified API. \hopsworks also comes with a Feature Store that enables
data scientists to write production-ready code, ensure data quality, and clean training data for ML models. \hopsworks allows users to start Jupyter Notebook servers where they can write applications in Python and PySpark, that can access services such as Spark, Hadoop Distributed File System (HDFS), and Flink in the cluster. Fig.~\ref{fig:hopsworks} shows the architecture of \hopsworks.
\subsection{The \hopsworks.ai cloud platform}
\hopsworks.ai~\cite{hopsworks.ai}, 
is a fully managed platform for running Hopsworks and the Feature Store in the cloud. It integrates seamlessly with third-party platforms such as Databricks, SageMaker, and KubeFlow. Users can install Hopsworks on the AWS cluster with the integration of Amazon EKS and Amazon ECR. Furthermore, users can also launch Hopsworks on Azure with the integration of Azure AKS and ACR.
\subsection{Project-based multi-tenancy on Hopsworks}
Hopsworks introduces the novel project-based multi-tenancy. A project in Hopsworks is a major unit of abstraction; it contains users, datasets(e.g., data and programs), services (e.g., Kafka, Jupyter Notebook servers), jobs, etc. Datasets are stored in HopsFS. A project is mapped on the HopsFS as a subdirectory in the \texttt{/Projects} directory. In Hopsworks, datasets of different projects or tenants are completely isolated. However, a dataset can be securely shared across different projects without necessarily replicating the dataset in HopsFS as part of the GDPR compliance. Hopsworks provides role-based access control within projects, with predefined “Data Owner” and ``Data Scientist'' roles, provided for GDPR compliance (``Data Owners'' are responsible for the data and access to the data, while ”Data Scientists” are processors of the data).

%% file: sections/design.tex
\section{Enabling Cloud-Native RStudio on Hopsworks: Design and Implementation}

This section describes the requirements, architecture proposal, and corresponding implementations of the system components that enable cloud-native RStudio on Hopsworks.
\subsection{System Requirements}
\label{requirements}

We have addressed the following functional and non-functional requirements when enabling cloud-native RStudio on Hopsworks. Functional requirements are as follows.
\begin{itemize}
    \item \textbf{On-demand RStudio server instances: } To avoid wasting resources, RStudio server instances should be launched or running only when the user needs them.
    \item \textbf{Easily launchable RStudio server instances: }It is important that the launching of RStudio server instances should not be complex and does not burden the users with difficult configuration tasks. Preferably, the launching of RStudio server instances should be initiated through a web interface.
    \item \textbf{Data sharing: }RStudio server instances should be able to read/write project data and programs(code) in HopsFS. In this way, users can collaborate on projects in their RStudio server instances without the need for external data-sharing tools outside of the Hopsworks ecosystem.
    \item \textbf{Logging: }Logging is an essential requirement for enterprise applications. Logging should be available on a tenant basis, and the logs should be stored for later use. 
\end{itemize}


In addition to the above-mentioned functional requirements, non-functional requirements to be met when enabling cloud-native RStudio on Hopsworks are discussed below.

\begin{itemize}
    \item \textbf{Isolation: }RStudio server instances of different users should be completely isolated. Though we would like data sharing with the same project, each user should be in control of their own instances, e.g., start, shutdown, and configuration. Furthermore, all traffic should be routed properly between RStudio server instances and clients running within the same project or different projects.  
    \item \textbf{Easy Deployment: }Deployment of RStudio should be automated and be integrated into the installation of Hopsworks, i.e., deployment of Hopsworks should include RStudio already installed.
    \item \textbf{Scalability: }The system should be able to scale horizontally by adding more nodes so that more RStudio servers can be run concurrently. This is essential for our use case for using our platform for teaching and learning R in a large class. Furthermore, the system should scale when users run data or compute-intensive workloads. It should therefore be possible to run RStudio jobs on a Spark cluster, e.g., Hopsworks cluster, Databricks cluster, etc.
    \item \textbf{Fault tolerance: }RStudio server instances may crash, but such cases should be independent (i.e., not affecting other instances) and not lead to any data loss.
    \item \textbf{Security: }All intra-communication between services should be encrypted with TLS. Data access control should be enforced when accessing data in HopsFS from RStudio. Furthermore, REST API access should always be authorized and validated.
\end{itemize}
\subsection{Architectural Overview}
\begin{figure}[!ht]
  \begin{center}
    \includegraphics[width=\columnwidth]{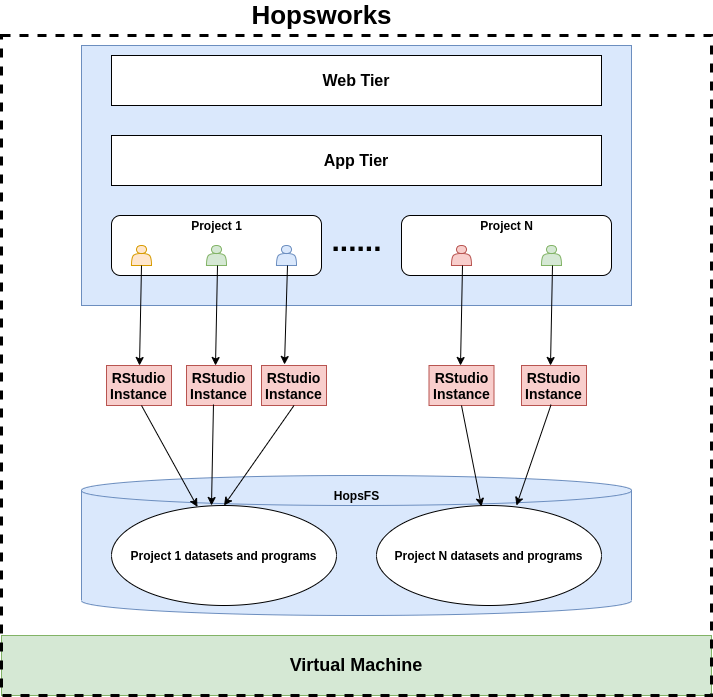}
  \end{center}
  \caption{System Model}
  \label{fig:system_model}
\end{figure}

Our model is on-demand \textit{single user RStudio server instances per project}. This model provides flexibility and better isolation because each user has control of their instance, e.g., start/stop their instance at a time they wish, and the failures in one server instance do not affect the other running instances. Furthermore, we can say there is better resource utilization because no resources are kept idle, i.e., an RStudio server instance runs only when a user needs it. The alternative models are one server instance per organization or one server instance per project. While we require single-user instances, users within the same project share data and programs. Figure \ref{fig:system_model} shows the system model. As shown in the figure, all users share the same web tier, application, and data layer on the same machine hence multi-tenancy. However, the project provides an abstraction of grouping some users to share some specific project resources.

\begin{figure*}
  \begin{center}
    \includegraphics[width=\textwidth]{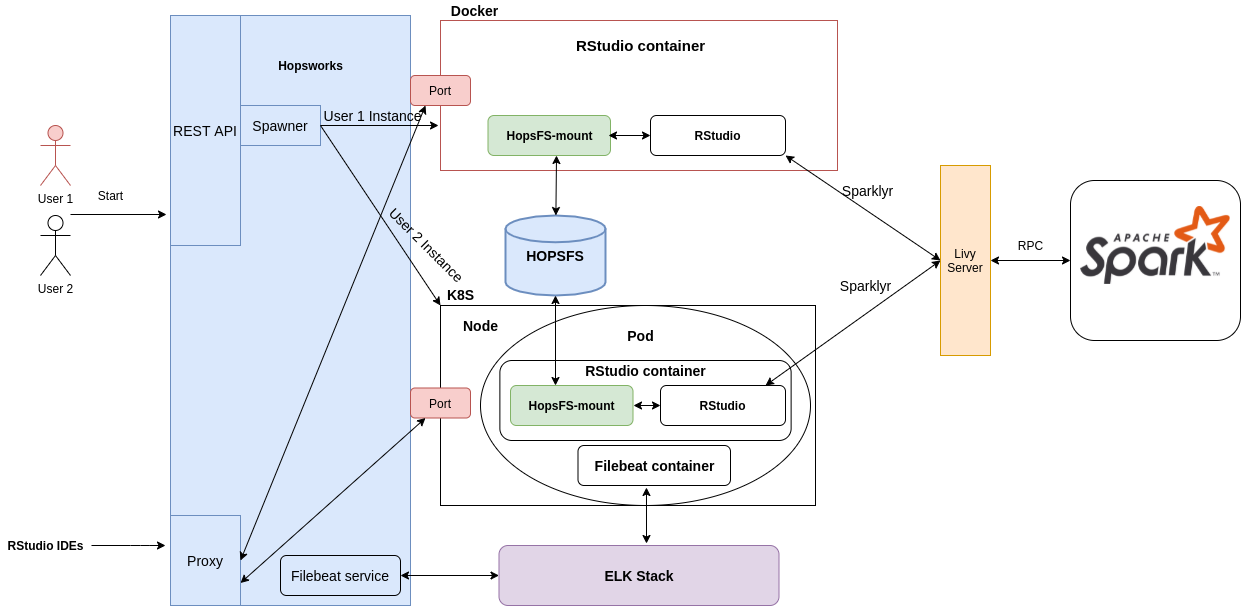}
  \end{center}
  \caption{System Architecture}
  \label{fig:system_archetecture}
\end{figure*}

Figure \ref{fig:system_archetecture} shows the system architecture and how the components interact with each other. The major components are REST API, Proxy, RStudio on Docker, RStudio on K8s, and HopsFS-mount. These will be discussed in the next section.
\subsection{System Components}
This section describes the major system components.

\subsubsection{RStudio REST API}
The RStudio API in Hopsworks has endpoints  that other services, like the WebService or front-end, can interact with for starting, stopping, or configuring RStudio servers within a project. All the REST endpoints require authentication and a valid project role.
\subsubsection{Reverse Proxy}
The role of the Reverse Proxy is to forward requests from the RStudio web UI to the appropriate RStudio server container and vice-versa. All the traffic is encrypted using TLS. The proxy server thus provides isolation of individual RStudio server instances by correctly redirecting traffic from the RStudio web UI to the correct RStudio server instance running as a microservice on Hopsworks and vice-versa.
\subsubsection{RStudio on Docker}
We use Docker to containerize RStudio, that is, to package the RStudio server, including all its dependencies (e.g., installing system-wide libraries), configuration, and start-up scripts. Our Docker image is built from the \textit{rocker/rstudio} \cite{rocker-rstudio} base image. The RStudio Docker containers are run as microservices on Hopsworks using the host network. When starting the container, a port where the RStudio server will listen is provided, and this port is exposed to the outside using the \texttt{EXPOSE} instruction. To separate containers of different users, the container is named in the following format: \texttt{\{project\_hdfs\_username\}\_rstudio}. The \texttt{\{project\_hdfs\_username\}} is unique for every user with a project. With this systematic naming of containers, a user is able to start a separate RStudio server instance in each project.
\subsubsection{RStudio on Kubernetes}
\begin{figure}[!ht]
  \begin{center}
    \includegraphics[width=\columnwidth]{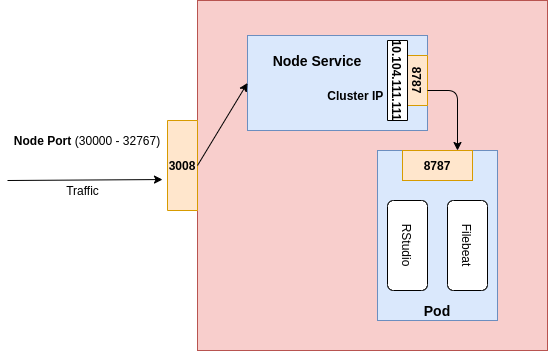}
  \end{center}
  \caption{Kubernetes System Architecture. The specified node port and cluster IP are examples.}
  \label{fig:rstudio_on_k8s_archtecture}
\end{figure}

In a Hopsworks deployment with Kubernetes, the RStudio servers are launched on Kubernetes. On the Hopsworks Kubernetes cluster, each project has its own namespace identified by the project name. In this project namespace, each user in the project has their own RStudio deployment, which has one replica set, two containers, a Nodeport service, ConfigMap, and Secrets. A deployment is named in the following format: \texttt{rstudio\_\{project\_name\}\_\{userid\}}. The containers in the deployment are the RStudio container and filebeat container. The filebeat container is for collecting the RStudio server logs. Each RStudio server container runs on port \texttt{8787}, but it is exposed to the outside network using the Nodeport service. The ConfigMap, which is immutable, is used to store the configuration files for the RStudio server and Sparklyr. The ConfigMap is mounted as a data volume in the Pod. The Secret is used to store the \texttt{jwt\_token} that is used by the container to authenticate with Hopsworks. Just like the ConfigMap, the Secret is also mounted as a volume in the Pod. Figure \ref{fig:rstudio_on_k8s_archtecture} shows the architecture on Kubernetes.
\subsubsection{Elasticsearch, Logstash, and Kibana (ELK) stack}
ELK stack is for aggregating, analyzing, and visualizing the logs from the RStudio application servers. The ELK stack comprises Elasticsearch, Logstash, and Kibana. Elasticsearch is a full-text search and analysis engine. Logstash is a log aggregator that collects data from various input sources. Kibana is a visualization layer that works on top of Elasticsearch, providing users with the ability to analyze and visualize the data. The logs are collected from the RStudio containers using the Filebeat service. The RStudio application writes the logs to a volume mounted in the container. When running RStudio on Docker, the Hopsworks Filebeat service collects the logs, while when running on Kubernetes, the Filebeat service is provided by the running container in the same Pod as the RStudio container. 
\subsubsection{HopsFS-mount}
In containerized applications, all data is lost once the container
is killed. However, we would like to persist all project resources (data, programs) after the user stops the RStudio server instance. One solution that Docker provides is \textit{volumes} that persist data on the host filesystem. One challenge with this is that some files can get very big (i.e., Big Data) and, therefore, can exceed the storage capacity of a single machine. 

To enable data sharing and persistent storage of data generated by users when using the RStudio server, we introduce \textit{HopsFS-mount}. HopsFS-mount allows us to mount the remote HopsFS in the RStudio container as a local filesystem. Once mounted, the user or application can operate on HopsFS using standard Unix utilities such as ls, cd, cp, mkdir, find, grep, etc. We then configure RStudio to read from and write to (user programs, data, packages) the mount point or directory, which in essence, means reading from and writing to HopsFS. Apart from enabling data persistence, storing data in HopsFS also enables resource sharing and collaboration among the project members.
\subsection{Authentication}
By default RStudio uses Linux Pluggable Authentication Module (PAM)~\cite{rstudio_authentication}\cite{pam}. This means that users log in using their Linux user account. To enable this, we create a Linux user each time the RStudio container is started and configure this user to be the user of the RStudio server running in the container. This Linux user is unique for each user within the project.
\subsection{Spark Integration}
We use Sparklyr~\cite{sparklyr}, an open-source package that provides an interface between R and Apache Spark. Sparklyr supports two deployment modes: local and cluster. To connect to the Hopsworks cluster, we use Apache Livy \cite{apache_livy}. Apache Livy is a service that runs on Hopsworks and enables easy interaction with a Spark cluster over a REST interface. With Apache Livy, Spark jobs or snippets of Spark code can be easily submitted to cluster and get execution results synchronously or asynchronously, all via a simple REST interface or an RPC client library.

Sparklyr provides various options for configuring both the behavior of the Sparklyr package as well as the underlying Spark cluster. We provide the Spark configuration to Sparklyr through a file named \textit{config.yml}, which is parsed and read into a \texttt{config} object required by Sparklyr. The configuration file is generated when the user is starting the RStudio server instance. The user can provide values to some of the fields in the configuration file through the Web UI. The fields include: \texttt{driverMemory},  \texttt{driverCores}, \texttt{executorMemory}, \texttt{executorCores} and \texttt{numExecutors}. Connecting through Livy also requires the URL to the Livy service and parameters for authentication. We have modified the Sparklyr package so that connecting to the Hopsworks cluster becomes easy for the user; the user needs to only put \texttt{method=``hopsworks''} in the \texttt{spark\_connect} method.  After the connection, the Livy Service starts the Spark context, \texttt{sc}, by requesting resources from the cluster manager and distributing work as usual. Since we modified the Sparklyr package, this version is installed as a system-wide library when building the Docker image.

\subsection{Package Management}
Packages are per project; that is, packages installed in Project A will not be visible in Project B. Packages in the RStudio server can be installed as system-wide libraries or user-specific packages. In our case, system-wide packages are installed when building the RStudio Docker images. User-specific packages are stored in HopsFS in the \texttt{DataSets/Rstudio/.Rpackages/\{hadoop\_username\}} path. The packages are accessible to RStudio through the HopsFS-mount and by setting the \texttt{R\_LIBS\_USER} variable in the packages directory in the mount point.

%% file: sections/discussion.tex
\section{Discussion}
This section summarizes how we addressed the challenges identified in our requirements (\ref{requirements}) for enabling cloud-native RStudio in a multi-tenant Hopsworks environment.

\subsection{Performance Isolation}
For each RStudio container running on Hopsworks, we set resource limits on how much memory and CPU it can use; thus, users who exceed their quota of allocated resources will not affect other users within the cluster. For containers not managed by Kubernetes, we use \textit{cgroups} to limit the memory and CPU the container can use, i.e., by passing \texttt{cpu-quota=<quota>} and \texttt{-m <memory>} flags in the  \texttt{docker run} command to specify the CPU quota and memory allocation respectively. For the containers orchestrated by Kubernetes, we set the resource limits in the \texttt{PodSpec} definition. These resource limits are variables that can be modified during cluster installation by the administrator.
\subsection{Security}
Each user within the project has their RStudio instance running in a separate Docker container. Docker provides security out of the box using Linux namespaces, control groups, and capabilities, among others, as earlier discussed. In the application layer, a role-based REST API with SSL termination is implemented. The Reverse proxy also properly proxies traffic from the RStudio server IDE to the RStudio server container. At the data layer, we securely mount HopsFS in the container such that each running container needs to provide SSL/TLS  certificates for each project to authenticate with HopsFS. Hopsworks provides other security features, including Single Sign On (SSO) for Active Directory, OAuth-2, and LDAP; data in transit encryption using two-way TLS and X.509 certificates; and role-based access control for project data.
\subsection{Data sharing}
In Hopsworks, datasets of different projects or tenants are completely isolated. However, a dataset can be securely shared across different projects without necessarily replicating the dataset in HopsFS as part of the GDPR compliance. Hopsworks provides role-based access control within projects, with predefined ``Data Owner'' and ``Data Scientist'' roles, provided for GDPR compliance (``Data Owners'' are responsible for the data and access to the data, while ``Data Scientists'' are processors of the data). We enable the secure mounting of HopsFS in the RStudio containers such that all the data security measures provided by Hopsworks are preserved. RStudio servers within the same project can then read and write project data in HopsFS.
\subsection{Scaling}
We use Kubernetes to launch
RStudio servers on the nodes. By adding more nodes to the system, we can therefore scale horizontally and launch more RStudio server instances. We have also extended the Sparklyr~\cite{sparklyr} R package so that users can launch Spark jobs in the Hopsworks Spark cluster over Livy to run data-intensive and compute-intensive workloads.

%% file: sections/evaluation-results.tex
\section{Cloud Resource Requirements and Scalability Evaluation}
We use Google Cloud Platform (GCP) as our IaaS provider to install Hopsworks and test our system in the cloud. Our Google Kubernetes Engine (GKE) cluster consisted of an n1-standard-16 head node (i.e., the control plane) and 4 n1-standard-8 worker nodes ~\cite{gcpspecs}. The n1-standard-16 has 16 cores and 60GB of RAM memory. Each n1-standard-8 node has 8 cores and 30GB of RAM memory. Hopsworks is installed on the head node.

The goal of the experiment was to test how many concurrent RStudio servers
can be run on our cluster before no more servers can be launched. To test the
system, 23 projects each with 2 users were created and each user in each project
launched an RStudio server. Each RStudio container has a limit of 2GB of
RAM that will be reserved by the Kubernetes scheduler when launching the container. The results showed that a total of 44 servers can be run concurrently on our cluster. Table~\ref{tab:cloud_perfoamnce_results} shows the projected resource requirements for a given number of RStudio servers. By adding more worker nodes to the cluster, the
number of concurrent RStudio servers can be increased.

\begin{table}[!ht]
    \caption{Projected memory and CPU requirements for given concurrent RStudio servers}
    \label{tab:cloud_perfoamnce_results}
\centering
\resizebox{\columnwidth}{!}{%
    \begin{tabular}{c|c|c} 
    \hline
      \textbf{\begin{tabular}[c]{@{}c@{}}Number of Concurrent\\  RStudio Servers\end{tabular}} & 
      \textbf{ Required RAM (GB)} & 
      \textbf{Min Num CPUs}\\ \hline
      10 & 30 & 8\\ \hline
      20 & 60 & 16\\ \hline
      40 & 120 & 32\\ \hline
    \end{tabular}
}
\end{table}

%% file: sections/conclusion.tex
\section{Conclusion and Future Work}
We have presented the design and implementation of integrating of RStudio on Hopsworks to enable a multi-user environment for RStudio users necessary for collaboration and resource sharing using cloud-native technologies. 
We solved the core challenges of performance isolation, security, and scaling endemic in a multi-tenant environment using this approach. Furthermore, we enabled secure data sharing (datasets and programs) among RStudio users on the platform, which is crucial for collaboration and data privacy. 

One interesting integration of RStudio within the Hopsworks ecosystem would be with the Hopsworks Feature Store, a data management system for managing machine learning features, including the feature engineering code and the feature data. Integrating RStudio, i.e., reading and writing data to Feature Store from RStudio, can allow easy management of machine learning features in R Machine Learning projects. Another improvement would be to enable running R scripts
as jobs in Hopsworks. 